\documentstyle[12pt,hyperref]{article}
\newcommand{\be}[1]{ \begin{equation}\label{#1} }
\newcommand{\ee}{\end{equation}}
\newcommand{\bea}[1]{\begin{eqnarray}\label{#1} }
\newcommand{\eea}{\end{eqnarray}}

\newcommand{\AAA}{{\cal A}}

\newcommand{\FF}{{\cal F}}
\newcommand{\NN}{{\cal N}}
\newcommand{\p}{\partial}
\newcommand{\wt}{\widetilde}
\def\ZZZ{{\hskip-3pt\hbox{ Z\kern-1.6mm Z}}}
\def\zzz{{\hskip-3pt\hbox{ z\kern-1mm z}}}

\def\ZZZ{{\hbox{Z\kern-1.6mm Z}}}
\def\zzz{{\hbox{z\kern-1mm z}}}

\newcommand{\BB}{{\cal B}}

\newcommand{\KK}{{\cal K}}

\newcommand{\HH}{{\cal H}}
\newcommand{\MM}{{\cal M}}

\newcommand{\EE}{{\cal E}}
\newcommand{\LL}{{\cal L}}

\newcommand{\wh}{\widehat}
\newcommand{\wc}{\check}

\newcommand{\wm}{\wc m}
\newcommand{\wn}{\wc n}
\newcommand{\wpp}{\wc p}
\newcommand{\wa}{\wc \alpha}
\newcommand{\wbb}{\wc \beta}
\newcommand{\wg}{\wc \gamma}

\newcommand{\ben}{\begin{eqnarray}\displaystyle}
\newcommand{\een}{\end{eqnarray}}
\newcommand{\refb}[1]{(\ref{#1})}
\newcommand{\sectiono}[1]{\section{#1}\setcounter{equation}{0}}

\def\one{{\hbox{ 1\kern-.8mm l}}}
\def\zero{{\hbox{ 0\kern-1.5mm 0}}}

\renewcommand{\HH}{\KK}
\renewcommand{\BB}{C}

\begin{document}

{}~
{}~

\hfill\vbox{\hbox{hep-th/0608182}}\break

\vskip .6cm

\medskip

\baselineskip 20pt 

\begin{center}

{\Large \bf
$\alpha'$-Corrections to Extremal Dyonic Black Holes
in Heterotic String Theory
}

\end{center}

\vskip .6cm
\medskip

\vspace*{4.0ex}

\centerline{\large \rm Bindusar Sahoo and Ashoke Sen}

\vspace*{4.0ex}

\centerline{\large \it Harish-Chandra Research Institute}

\centerline{\large \it  Chhatnag Road, Jhusi,
Allahabad 211019, INDIA}

\vspace*{1.0ex}

\centerline{E-mail: bindusar@mri.ernet.in, ashoke.sen@cern.ch,
sen@mri.ernet.in}

\vspace*{5.0ex}

\centerline{\bf Abstract} \bigskip

We explicitly compute the entropy of an extremal dyonic black hole in
heterotic string theory compactified on $T^6$ or $K3\times T^2$
by taking into account all the tree level four derivative corrections
to the low energy effective action. For supersymmetric black holes
the result agrees with the answer obtained earlier 1) by including only
the Gauss-Bonnet corrections to the effective
action  2) by including all terms related to the
curvature squared terms via space-time supersymmetry transformation, and
3) by using general arguments based on the assumption of $AdS_3$
near horizon geometry and space-time supersymmetry.
For non-supersymmetric extremal black holes the result agrees
with the one based on the assumption of $AdS_3$
near horizon geometry and space-time supersymmetry of the underlying 
theory.

\vfill \eject

\baselineskip 18pt

\tableofcontents

\sectiono{Introduction}

String theory at low energy describes Einstein gravity coupled
to certain matter fields, together
with infinite number of higher derivative corrections. Thus study
of black holes in string theory involves study of black holes in
higher derivative theories of gravity. While this is a complicated
problem for general black holes, there are various techniques 
available for studying higher derivative corrections to the entropy
of extremal black holes with or without supersymmetry. Nevertheless
most of the analysis so far has been done by taking into account
only a subset of these corrections, {\it e.g.} by including only the
terms in the action proportional to 
Gauss-Bonnet term\cite{9711053}, or by
including the set of all terms which are related to the curvature squared
terms by supersymmetry 
transformation\cite{9801081,9812082,9904005,9906094,
9910179,0007195,0009234,0012232}\footnote{See 
\cite{0506251} for some
discussion on the relation between these two approaches.}. 
Even at the string tree level there
are other four derivative terms in the action which are {\it a
priori} equally important, and hence there is no justification for
not including these terms in the analysis. Later 
refs.\cite{0506176,0508218} proved certain non-renormalization
theorems establishing that for a certain class of
supersymmetric black holes the
results of \cite{9711053, 9812082,9904005,9906094,
9910179,0007195,0009234,0012232} are in fact exact.
The underlying assumption behind this proof is the existence of an
$AdS_3$ component of the near horizon geometry of the black hole
solution when embedded in the full ten dimensional space-time, and
supersymmetry of the resulting two dimensional theory that lives on
the boundary of this $AdS_3$.

Notwithstanding these non-renormalization theorems, it is important
to verify the result by a direct calculation that takes into account all
the higher derivative corrections in a given order. An attempt in this
direction was made in \cite{0607094} where the author tried to
include all the tree level four derivative corrections to the
action of heterotic string theory compactified on a six dimensional
torus $T^6$, and used this to compute correction to the entropy of
an extremal dyonic black hole\cite{9507090}. 
The apparent conclusion of this paper was that 
the entropy computed this way disagrees with the earlier results based
on the calculations of \cite{9711053,9801081,9812082,
9904005,9906094,9910179,0007195,0009234,0012232}. 
If this is correct then
this would also contradict the non-renormalization theorems of
\cite{0506176,0508218}. 
A closer look however reveals that the analysis
of \cite{0607094} left out one important term, -- the coupling
of the gravitational Chern-Simons term to the 3-form field strength. 

The
purpose of this paper is to recalculate the entropy of a dyonic black hole
in tree level heterotic
string theory by including the complete set of tree level
four derivative terms in the heterotic string effective action. We find that
after the effect of gravitational Chern-Simons term is included, the
resulting entropy agrees perfectly with the results of earlier analysis,
in accordance with the non-renormalization 
theorems of \cite{0506176,0508218}.

In carrying out our analysis we  use the entropy function 
formalism\cite{0506177}
which is well suited for studying higher derivative 
corrections\cite{9307038,9312023,9403028,9502009} to the
entropy of extremal black 
holes.  In the specific context of heterotic string
theory in four dimensions, this formalism has been used to calculate
the extremal black hole entropy in the presence of
Gauss-Bonnet term\cite{0508042}, as well as in the presence
of all terms related to the curvature squared terms via space-time
supersymmetry transformaion\cite{0603149}.  
It was also used in the analysis of \cite{0607094} for computing the
effect of all the four derivative terms at tree level heterotic string theory
except the gravitational Chern-Simons term. In general the 
computation of the entropy function
involves expressing the four dimensional Lagrangian density
in a fully gauge and
general covariant form involving only the gauge field strengths, metric,
Riemann tensor, scalar fields and
their covariant derivatives, and then evaluating it in a generic
$SO(2,1)\times SO(3)$  invariant background reflecting the isometry of
the $AdS_2\times S^2$ near horizon geometry of an extremal black hole.
For part of the four dimensional lagrangian density which comes from
the dimensional reduction of a manifestly covariant six
dimensional lagrangian density, the
contribution to the entropy function
can be related to the value of the
six dimensional Lagrangian density evaluated in the
corresponding six dimensional background\cite{0607094}. 
This avoids the necessity
of first dimensionally reducing the six dimensional lagrangian density
to four
dimensions and then evaluating its value. However this procedure fails
for a part of the six dimensional lagrangian density
that involves the gravitational Chern-Simons term coupled to
the 3-form field strength, since this term
cannot be written in a manifestly covariant form.
Thus we need to first dimensionally  
reduce this term 
to four dimensions, express it in a manifestly covariant
form after throwing away total derivative terms 
and then evaluate its value in a specific
background geometry. A general procedure for dealing with
dimensionally reduced Chern-Simons terms in the entropy function
formalism was developed in \cite{0601228}. Thus the entropy function
formalism is well-suited for studying the problem at hand.

In section \ref{s1} we discuss the general strategy for dealing with the
dimensional reduction of a six dimensional action that contains a
gravitational Chern-Simons term in the definition of the 3-form
field strength. We also discuss the strategy for computing 
the entropy function in such a theory. 
In section \ref{s2} we consider the specific example
of tree level heterotic string theory compactified on 
$T^6$ or $K3\times T^2$, analyze the complete low 
energy effective action up to 4-derivative terms and 
evaluate its contribution
to the entropy function. The extremal black hole entropy, 
given by the
value of the entropy function at its extremum, is then
shown to match the
results of the earlier computation of 
\cite{9711053,9801081,9812082,9904005,9906094,
9910179,0007195,0009234,0012232,0506177,
0508042,0603149}
based on only 
a subset of the 4-derivative
corrections to the Lagrangian density.

\sectiono{Strategy for Dealing with Chern-Simons Terms}\label{s1}

We begin with the low energy effective field theory
of ten dimensional heterotic 
string theory compactified
on $T^4$ or $K3$. At tree level
there is a consistent truncation of this theory
in which we ignore all the ten dimensional gauge fields
and the massless fields 
associated with the components
of the metric and the anti-symmetric tensor fields along the 
compact space $T^4$ or $K3$. In this case the remaining massless
fields consist of the string metric $G^{(6)}_{MN}$, the anti-symmetric
tensor field $B^{(6)}_{MN}$ and the dilaton field $\Phi^{(6)}$ with 
$0\le M,N\le 5$. The gauge invariant field strength associated
with the anti-symmetric tensor field is given by:
\be{e1}
H^{(6)}_{MNP}=\p_{M} B^{(6)}_{NP} 
+ \p_{N} B^{(6)}_{PM} + \p_{P} B^{(6)}_{MN}
+ \lambda \, \Omega^{(6)}_{MNP}\, ,
\ee
where $\lambda$ is a coefficient to be specified later and 
$\Omega^{(6)}_{MNP}$ denotes the 
gravitational 
Chern-Simons 3-form constructed out of the six dimensional
spin connections, normalized such that 
\bea{ecsnorm}
&&
\p_Q \Omega^{(6)}_{MNP} + \hbox{anti-symmetrization
in $P,Q,M,N$}
\cr
&=& -{1\over 8} \, R^{(6)K}_{~~~~SMN} \, R^{(6)S}_{~~~~KPQ}
+ \hbox{anti-symmetrization
in  $P,Q,M,N$}\, .
\eea
$R^{(6)}_{MNPQ}$ denotes the Riemann tensor associated with the
metric $G^{(6)}_{MN}$.
We shall denote
the action of this theory as
\be{e2}
S = \int d^6 x \, \sqrt{-\det G^{(6)}} \, \LL^{(6)}
\ee
where 
the Lagrangian density $\LL^{(6)}$ 
is a function of $G^{(6)}_{MN}$,
the Riemann tensor $R^{(6)}_{MNPQ}$, 
$H^{(6)}_{MNP}$, $\Phi^{(6)}$ and covariant derivatives
of these fields.

We shall study compactification of this theory on a two dimensional
torus $T^2$ and study the entropy of extremal black holes in this
theory. 
This will give rise to four abelian gauge fields from the
components of the metric and the antisymmetric tensor 
fields along
the $T^2$ directions.
The resulting lagrangian density, besides depending on the
covariant objects like the metric, Riemann tensor, gauge field strengths
and their covariant derivatives, 
will also depend explicitly on the spin connection and the gauge fields
due to the presence of the gravitational  Chern-Simons term 
inside $H_{MNP}$
as in \refb{e1} and similar gauge Chern-Simons terms which are
induced during compactification\cite{9207016}.
Our goal is to
express the effective Lagrangian density 
in a manifestly covariant form
without involving any Chern-Simons terms so that we can apply the
entropy function formalism. This will be done in two steps:

\begin{enumerate}

\item First at the level of the six dimensional description itself
we shall introduce a new field $\BB^{(6)}_{MN}$ and its field 
strength
\be{e3}
\HH^{(6)}_{MNP} = \p_{M} \BB^{(6)}_{NP}
+ \p_{N} \BB^{(6)}_{PM} + \p_{P} \BB^{(6)}_{MN}\, ,
\ee
and consider a new Lagrangian density
\bea{e4}
\sqrt{-\det G^{(6)}}\, \wt\LL^{(6)} &\equiv& 
\sqrt{-\det G^{(6)}}\, 
\LL^{(6)} + {1\over 16\pi^2} {1\over (3!)^2}\, 
\epsilon^{MNPQRS} \HH^{(6)}_{MNP} H^{(6)}_{QRS} 
\nonumber \\
&& - {1\over 16\pi^2} {1\over (3!)^2}\, 
\lambda \, \epsilon^{MNPQRS} \HH^{(6)}_{MNP} 
\Omega^{(6)}_{QRS}\, 
\eea
where we 
treat $H^{(6)}_{MNP}$ and $\BB^{(6)}_{MN}$ as independent
variables. The normalization factor of 
${1\over 16\pi^2} {1\over (3!)^2}$ has been introduced for later
convenience.
Then we can first solve the $\BB^{(6)}_{MN}$ 
equations of motion
to get the result
\be{e5}
d (H^{(6)}-\lambda \, \Omega^{(6)}) = 0\, ,
\ee
which can then be solved to get \refb{e1}. Substituting this into
\refb{e4} we recover the original action
\refb{e2}. On the other hand if we first eliminate 
$H^{(6)}_{MNP}$ by using its equation of motion, we get
\be{e6}
\sqrt{-\det G^{(6)}}\, \wt\LL^{(6)} = \sqrt{-\det G^{(6)}}\, 
\wt\LL^{(6)\prime} 
- {1\over 16\pi^2} {1\over (3!)^2}
\lambda \, \epsilon^{MNPQRS} \HH^{(6)}_{MNP} 
\Omega^{(6)}_{QRS}\,
\ee
where $\wt\LL^{(6)\prime}$
is the sum of the first two terms on the 
right hand
side of \refb{e4} after elimination of 
$H^{(6)}_{MNP}$. This is now to be regarded
as a function of the `dual field' $\BB^{(6)}_{MN}$. 
$\wt\LL^{(6)\prime}$ depends on $\BB^{(6)}_{MN}$
solely through its field strength $\HH^{(6)}\propto
d\BB^{(6)}$ 
and hence has a manifestly
covariant form without any Chern-Simons terms.
The full Lagrangian density is still not manifestly
covariant due to the presence of the Chern-Simons
3-form in the last term of \refb{e6}.

\item We now dimensionally reduce this theory to four 
dimensions by
introducing the fields $G_{\mu\nu}$, $\BB_{\mu\nu}$, $\Phi$, 
$\wh G_{mn}$, $\wh \BB_{mn}$
and $\AAA_\mu^{(i)}$ ($0\le\mu\le 3$,   $4\le m,n\le 5$,
$1\le i\le 4$) via the
relations
\bea{eag2}
&& \wh G_{mn} = G^{(6)}_{mn}, 
\quad \wh \BB_{mn} = \BB^{(6)}_{mn}\, , 
\nonumber 
\\  && \wh G^{mn} = (\wh G^{-1})^{mn} \, , \nonumber  \\
&& \AAA^{(m-3)}_\mu = {1\over 2} 
\wh G^{mn} G^{(6)}_{m\mu} , \quad
\AAA^{(m-1)}_\mu = {1\over 2} \BB^{(6)}_{m\mu} - \wh \BB_{mn} 
\AAA^{(n-3)}_\mu \, , \nonumber \\
&& G_{\mu\nu} = G^{(6)}_{\mu\nu} - \wh G^{mn} G^{(6)}_{m\mu} 
G^{(6)}_{n\nu}\, , \nonumber \\
&& \BB_{\mu\nu} = \BB^{(6)}_{\mu\nu} 
- 4 \wh \BB_{mn} \AAA^{(m-3)}_\mu \AAA^{(n-3)}_\nu - 2 
(\AAA^{(m-3)}_\mu \AAA^{(m-1)}_\nu
- \AAA^{(m-3)}_\nu \AAA^{(m-1)}_\mu) \, , \nonumber \\
&& \Phi = \Phi^{(6)}  
-{1\over 2} \ln V_\MM\, ,
\een
where $x^4$ and $x^5$ are the coordinates labelling the torus and
$V_\MM$ is the volume of $T^2$ measured in the string
metric. We shall normalize $x^4$ and $x^5$ so that they have
coordinate radius $\sqrt{\alpha'}=4$.
Then
\be{evm}
V_\MM= 64\pi^2 \sqrt{\det \wh G}\, .
\ee
The gauge invariant field strengths associated with 
$\AAA_\mu^{(i)}$ and $\BB_{\mu\nu}$ are
\be{eag2a}
\FF^{(i)}_{\mu\nu} = \p_\mu \AAA^{(i)}_\nu - 
\p_\nu \AAA^{(i)}_\mu\, , 
\qquad 1\le i,j\le 4\, ,
\ee
\be{eag2b}
\HH_{\mu\nu\rho} = (\p_\mu \BB_{\nu\rho} + 2 
\AAA_\mu^{(i)} L_{ij} 
\FF^{(j)}_{\nu\rho}) 
+ \hbox{cyclic permutations of $\mu$, $\nu$, $\rho$}\, , 
\ee
where
\be{edefl}
L = \pmatrix{ 0 & I_2\cr I_2 & 0}\, ,
\ee
$I_2$ being $2\times 2$ identity matrix.
In this case the Lagrangian density, obtained by dimensional
reduction of the right hand side
of \refb{e6} has the form
\be{ed0}
\sqrt{-\det G} \, \wt\LL = \sqrt{-\det G} \, \wt\LL' 
+ \sqrt{-\det G} \, \wt\LL''\, ,
\ee
where
\be{ed1a}
\sqrt{-\det G} \, \wt\LL'=
\int dx^4 dx^5\, \sqrt{-\det G^{(6)}}\, 
\wt\LL^{(6)\prime} 
\, ,
\ee
\be{ed1b}
\sqrt{-\det G} \, \wt\LL'' =  - {1\over 16\pi^2} {1\over (3!)^2}
\lambda \, \int dx^4 dx^5 \,
\epsilon^{MNPQRS} \HH^{(6)}_{MNP} 
\Omega^{(6)}_{QRS} + \hbox{total derivative terms}\, .
\ee
$\wt\LL'$ is a function of the field 
strength $\HH_{\mu\nu\rho}$
and other covariant objects. 
We shall explicitly
demonstrate that $\wt\LL''$ is also a function of the field strengths
and other covariant objects after we remove certain
total derivative terms.
However due to
the presence of explicit gauge
fields in the expression for $\HH_{\mu\nu\rho}$ this form of the
Lagrangian density 
is not suitable for applying the entropy function method.
For this we dualize this action further by replacing the Lagrangian
density $\sqrt{-\det G}\, \wt\LL$ by
\be{ed2}
\sqrt{-\det G}\, 
\wt\LL + \epsilon^{\mu\nu\rho\sigma} \, \HH_{\mu\nu\rho} \,
\p_\sigma \, b +  3  \, b\, \epsilon^{\mu\nu\rho\sigma}\,  
\FF_{\sigma\mu}^{(i)} L_{ij} 
\FF^{(j)}_{\nu\rho}\, ,
\ee
and treating $\HH_{\mu\nu\rho}$ and the new scalar field $b$ as
independent variables. If we choose to first use the 
equation of motion of the $b$ field then we get 
\be{ed3}
\epsilon^{\mu\nu\rho\sigma} \p_\sigma
\left( \HH_{\mu\nu\rho} - 6 \AAA_\mu^{(i)} L_{ij} 
\FF^{(j)}_{\nu\rho} \right) = 0\, ,
\ee
which has as its solution the form \refb{eag2b} 
for some $\BB_{\mu\nu}$. Substituting this into \refb{ed2}
we recover the original action \refb{ed0} up to total derivative terms.  
On the other hand if
we first 
eliminate $\HH_{\mu\nu\rho}$ from \refb{ed2}
by its equation of motion
we shall get a Lagrangian density of the form:
\be{ed4}
\sqrt{-\det G} \, \wh\LL = \sqrt{-\det G} \, \wh\LL'
+  3  \, b\, \epsilon^{\mu\nu\rho\sigma}\,  
\FF_{\sigma\mu}^{(i)} L_{ij} 
\FF^{(j)}_{\nu\rho}\, ,
\ee
where $\wh\LL'$, obtained by substituting the
solution for $\HH_{\mu\nu\rho}$ in the first two terms in
\refb{ed2}, has a manifestly covariant expression in terms
of
$\p_\sigma b$
and other covariant objects.  
This way we arrive at a manifestly covariant form
of the Lagrangian density for which we can apply the entropy
function formalism.

\end{enumerate}

Let us now say a few words about the evaluation of the
entropy function
$\EE$.  For this we need to consider a general $AdS_2\times S^2$
near
horizon geometry with all other
background field configurations
consistent with the symmetries of $AdS_2\times S^2$ and define
\be{ent1}
\EE = 2\pi \left( \sum_{i=1}^4 \wt q_i \,
\wt e_i - \int d\theta \, d\phi \,
\sqrt{-\det G} \, \wh\LL\right)\, ,
\ee
evaluated in this background.
Here $\wt q_i$ denotes the electric charge associated with the gauge
field $\AAA_\mu^{(i)}$ and $\wt e_i$ denotes the value of the radial
electric field $\FF^{(i)}_{rt}$.
Since \refb{ed4} is obtained from \refb{ed2} after elimination
of the variables $\HH_{\mu\nu\rho}$, and since the right hand side of
\refb{ed2} is manifestly covariant when
$\HH_{\mu\nu\rho}$ is interpreted as an auxiliary field, we 
can replace the $\sqrt{-\det G} \, \wh\LL$ on the right
hand side of \refb{ent1} by the right hand side of \refb{ed2}. 
Since both
$\p_\sigma b$ and $\HH_{\mu\nu\rho}$ vanish in an
$AdS_2\times S^2$
geometry due to the absence of $SO(2,1)\times SO(3)$ invariant
1- and 3-forms, we can set them to zero in \refb{ed2} during the
computation of the entropy function. Thus we have
\be{ent2}
\EE = 2\pi \left( \sum_{i=1}^4 \wt q_i \, \wt e_i - \int d\theta 
\, d\phi \,
\sqrt{-\det G}\, 
\wt\LL -    3  \int d\theta d\phi
\, b\, \epsilon^{\mu\nu\rho\sigma}\,  
\FF_{\sigma\mu}^{(i)} L_{ij} 
\FF^{(j)}_{\nu\rho} \right)\, .
\ee
Using eqs.\refb{ed0}, \refb{ed1a} we can express this as
\bea{ent3}
\EE &=& 2\pi \bigg( \sum_{i=1}^4 \wt q_i \, \wt e_i
- \int d\theta  \, d\phi  \, dx^4  \, dx^5\, \sqrt{-\det G^{(6)}}\, 
\wt\LL^{(6)\prime}
 - \int d\theta \, d\phi\, 
\sqrt{-\det G}\, 
\wt\LL'' \nonumber \\
&& -    3  \int d\theta \,  d\phi
\, b\, \epsilon^{\mu\nu\rho\sigma}\,  
\FF_{\sigma\mu}^{(i)} \, L_{ij} \,
\FF^{(j)}_{\nu\rho} \bigg)\, .
\eea
Finally, using \refb{e4}, \refb{e6} we can express this as
\bea{ent4}
\EE &=& 2\pi \Bigg[ \sum_{i=1}^4 \wt q_i   \wt e_i
- \int d\theta   d\phi  dx^4   dx^5\,  \left(
\sqrt{-\det G^{(6)}}\, 
\LL^{(6)} + {1\over 16\pi^2} {1\over (3!)^2}\, 
\epsilon^{MNPQRS} \HH^{(6)}_{MNP} H^{(6)}_{QRS} 
\right) \nonumber \\
&&- \int d\theta \, d\phi \,
\sqrt{-\det G}\, 
\wt\LL''  -    3  \int d\theta \, d\phi
\, b\, \epsilon^{\mu\nu\rho\sigma}\,  
\FF_{\sigma\mu}^{(i)} L_{ij} 
\FF^{(j)}_{\nu\rho} \Bigg]\, ,
\eea
where $H^{(6)}_{MNP}$ needs to be interpreted as an elementary
auxiliary field which has to be eliminated by its equation of
motion.
The terms in the first line of \refb{ent4} can be evaluated by 
regarding the background as a six dimensional
configuration. Thus we do not need to explicitly find the 
dimensional reduction of this term.
For the contribution from the $\wt\LL''$ term
however we cannot directly evaluate the six dimensional
form proportional to $\int dx^4 dx^5 \,
\epsilon^{MNPQRS} \HH^{(6)}_{MNP} 
\Omega^{(6)}_{QRS}$ due to the presence of the total derivative terms
in \refb{ed1b}. We need to first find its dimensional reduction to
four dimensions and then use this to calculate the entropy function.

So far  we have not made any approximation.
What we are interested in however is an approximation scheme
where we take into account higher derivative corrections to
the effective action in a power series expansion. In particular
we shall be interested in the correction due to the
four derivative terms in the action. For this let us split the
original Lagrangian density $\LL^{(6)}$ as 
\be{eap1}
\LL^{(6)}=\LL^{(6)}_0+\LL^{(6)}_1\, ,
\ee
where $\LL^{(6)}_0$ denotes the supergravity Lagrangian
density and 
$\LL^{(6)}_1$ denotes four derivative corrections. 
The entropy function obtained from this Lagrangian density 
has the form:
\be{eap4}
\EE=\EE_0+\EE_1\, ,
\ee
with $\EE_0$ and $\EE_1$ reflecting the contribution from the two
and four derivative terms respectively:
\bea{ee0}
\EE_0 &=& 2\pi \bigg( \sum_{i=1}^4 \wt q_i \wt e_i
- \int d\theta d\phi dx^4 dx^5\,  \left(
\sqrt{-\det G^{(6)}}\, 
\LL^{(6)}_0 + {1\over 16\pi^2} {1\over (3!)^2}\, 
\epsilon^{MNPQRS} \HH^{(6)}_{MNP} H^{(6)}_{QRS} 
\right) \nonumber \\
&& -    3  \int d\theta d\phi
\, b\, \epsilon^{\mu\nu\rho\sigma}\,  
\FF_{\sigma\mu}^{(i)} L_{ij} 
\FF^{(j)}_{\nu\rho} \bigg)\, ,
\eea
\be{ee1}
\EE_1= 2\pi \bigg( - \int d\theta d\phi dx^4 dx^5\,  
\sqrt{-\det G^{(6)}}\, 
\LL^{(6)}_1
- \int d\theta \, d\phi \, 
\sqrt{-\det G}\, 
\wt\LL'' 
\bigg)\, .
\ee
Since the entropy is given by the value of $\EE$ at its {\it extremum},
a first order error in the determination of the near horizon background
will give a second order error in the value of the entropy. Thus we
can find the near
horizon background, including the auxiliary field
$H^{(6)}_{MNP}$,
by extremizing $\EE_0$ and then evaluate 
$\EE_0+\EE_1$ in this background. This
gives the value of the entropy correctly up to first
order.

\sectiono{Computation of the Entropy} \label{s2}

We shall now compute 
the entropy function for
heterotic string theory compactified on $T^6$ or $K3\times T^2$
following the strategy outlined in the previous section. We begin with
the computation of $\EE_0$. In the $\alpha'=16$ unit that we shall be
using in order to facilitate comparison with previous results
({\it e.g} that of \cite{0508042}), the relevant 
bosonic part of the Lagrangian 
density 
$\LL^{(6)}_0$, describing heterotic string theory compactified
on $T^4$ or $K3$, can be expressed as
\be{e2.1}
\LL^{(6)}_0 = {1\over 32\pi}\,  e^{- 2 \Phi^{(6)}}\, 
\left[ R^{(6)} + 4 \p_M\Phi^{(6)} \p^M\Phi^{(6)} 
- {1\over 12} \,
H^{(6)}_{MNP} H^{(6)MNP}\right] \, ,
\ee
where all the indices are raised and lowered 
by the six dimensional string metric $G^{(6)}_{MN}$.
In writing down this expression we have set to zero all the ten
dimensional gauge fields as well as the gauge and moduli fields
associated with the compact space $T^4$ or $K3$. This is a consistent
truncation of the theory.
Thus at this order $H^{(6)}_{MNP}$, obtained by extremizing
$\EE_0$ given in \refb{ee0},
is given
by 
\be{e2.3}
H^{(6)MNP} = -{1\over 3!}\, {2\over\pi}\, 
(\sqrt{-\det G^{(6)}})^{-1}\, e^{2\Phi^{(6)}}\,
\epsilon^{MNPQRS} \HH^{(6)}_{QRS}
\, .
\ee
As discussed after eq.\refb{ee1}, we can continue to use this result even 
at next order if we want to calculate the correction to the black hole 
entropy up to four derivative terms.

After dimensional reduction given in \refb{eag2} we get a
four dimensional theory. 
We consider an extremal black hole solution in this theory
with
near horizon configuration:
\bea{e2.4}
&& ds^2 \equiv G_{\mu\nu} dx^\mu dx^\nu 
= v_1\left(-r^2 dt^2 + {dr^2\over r^2}\right) +
v_2(d\theta^2 + \sin^2\theta d\phi^2)\, , \nonumber \\
&& \wh G = \pmatrix{ u_1^2 & 0\cr 
0 & u_2^2}\, , \qquad \wh \BB=0, \qquad
\qquad e^{-2\Phi}
=u_S\, , \qquad b = 0\, ,
\nonumber \\
&& \FF^{(1)}_{rt} = \wt e_1, \qquad \FF^{(3)}_{rt} 
= {1\over 16}\, \wt e_3, \qquad
\FF^{(2)}_{\theta\phi} = {\wt p_2\over 4\pi} \sin\theta\, , \qquad
\FF^{(4)}_{\theta\phi} = {\wt p_4\over 64\pi} \sin\theta\, ,
\eea
where an extra factor of $1/16$ has been included in the expressions
for $\FF^{(3)}_{rt}$ and $\FF^{(4)}_{\theta\phi}$ for later
convenience.
We have set the off-diagonal components of $\wh G$, $\wh\BB$, the
scalar field $b$ and some components of the eletromagnetic field
strengths to zero by requiring the field configuration to be invariant
under $x^5\to -x^5$, $x^i\to -x^i$ for $1\le i\le 3$.
Using \refb{eag2} we see that this corresponds to 
the following six
dimensional field configuration:
\bea{e2.5}
&& ds_6^2 \equiv G^{(6)}_{MN} dx^M dx^N
= ds^2 + u_1^2 ( dx^4+ 2 \wt e_1 r 
dt)^2 +
u_2^2 \left(dx^5 - {\wt p_2\over 2\pi} \cos\theta d\phi\right)^2
\, ,\nonumber \\
&&  
\BB^{(6)}_{4t}  = {1\over 8} \wt e_3 r \, , \qquad
\BB^{(6)}_{5\phi} = 
-{\wt p_4\over 32\pi} \cos\theta  \, ,
\nonumber \\
&& e^{-2\Phi^{(6)}} =  {u_S\over 64\pi^2\, u_1 u_2}\, ,
\eea
which gives
\be{e2.5a}
 \HH^{(6)}_{rt4}   = 
-{1\over 8}\,  \wt e_3 \, , \qquad \HH^{(6)}_{\theta \phi 5} = 
-{\wt p_4\over 32\pi} \sin\theta  \, .
\ee
We shall use the convention
\be{e2.6}
\epsilon^{tr\theta\phi 45} = 1\, .
\ee
Eq.\refb{e2.3} then gives
\bea{e2.7}
H^{(6)rt4} &=& {2\over \pi}\,
(\sqrt{-\det G^{(6)}})^{-1}\, e^{2\Phi^{(6)}}\,
\HH^{(6)}_{\theta\phi 5} = -{4\over v_1 v_2 u_S}\,
{\wt p_4}, \nonumber \\
H^{(6)\theta\phi 5} &=& -{2\over \pi}\,
(\sqrt{-\det G^{(6)}})^{-1}
\, e^{2\Phi^{(6)}}\,\HH^{(6)}_{rt4} = {16\pi
\over v_1 v_2 u_S\sin\theta}\wt e_3\, .
\eea
For this specific configuration \refb{ee0} gives the leading
order entropy function to be
\be{e2.8}
\EE_0 = 2\pi \left[ \wt e_1 \wt q_1 + \wt e_3 \wt q_3
-{1\over 8} v_1 v_2 u_S \left( -{2\over v_1} + {2\over v_2}
+{2 u_1^2 \wt e_1^2\over v_1^2} +{128\pi^2 u_2^2 \wt e_3^2
\over v_1^2 u_S^2}- {u_2^2 \wt
p_2^2\over 8\pi^2 v_2^2}
- {8 u_1^2 \wt p_4^2 \over v_2^2 u_S^2} \right)\right]\, .
\ee
Extremizing this with respect to $\wt e_1$ and $\wt e_3$ and
substituting their values back in \refb{e2.8} we get
\be{e2.9}
\wt e_1 = {2 v_1 \wt q_1\over v_2 u_S u_1^2}\, ,
\qquad \wt e_3 = {v_1 u_S \wt q_3\over 32\pi^2 v_2 u_2^2}\, .
\ee
and 
\be{e2.10}
\EE_0 = {\pi \over 4} v_1 v_2 u_S \left[{2\over v_1} - {2\over v_2}
+ {8 \wt q_1^2 \over v_2^2 u_S^2 u_1^2}
+ {\wt q_3^2 \over 8\pi^2 v_2^2 u_2^2}+ 
{u_2^2 \wt
p_2^2\over 8\pi^2 v_2^2}
+ {8 u_1^2 \wt p_4^2 \over v_2^2 u_S^2}\right]\, .
\ee

In this form the entropy function cannot be directly compared with the 
earlier results of \cite{0508042}, since
we have defined 
the gauge
fields $\AAA_\mu^{(3)}$ and $\AAA_\mu^{(4)}$ 
via dimensional reduction of 
the fields 
$\BB^{(6)}_{MN}$ 
whereas the gauge fields $A_\mu^{(3)}$ and $A_\mu^{(4)}$ of
ref.\cite{0508042} would
come from the dimensional reduction of the 
anti-symmetric tensor field $B^{(6)}_{MN}$ which
are dual to the fields $\BB^{(6)}_{MN}$. 
We can find the relation between the charges $(\wt p_i,\wt q_i)$ and the 
charges $(p_i,q_i)$ of \cite{0508042}  by comparing the
expressions for $H^{(6)MNP}$ given in \refb{e2.7} with the
corresponding expressions in \cite{0508042}, and then using the 
relation between the near horizon fields and charges in both description. 
This gives
\be{e2.11}
q_1=\wt q_1, \quad p_2 =\wt p_2, \quad q_3 = -\wt p_4, \quad
p_4 = -\wt q_3\, .
\ee
\refb{e2.10} may now be rewritten as
\be{e2.10a}
\EE_0 = {\pi \over 4} v_1 v_2 u_S 
\left[{2\over v_1} - {2\over v_2}
+ {8  q_1^2 \over v_2^2 u_S^2 u_1^2}
+ {p_4^2 \over 8\pi^2 v_2^2 u_2^2}+ 
{u_2^2  
p_2^2\over 8\pi^2 v_2^2}
+ {8 u_1^2 q_3^2 \over v_2^2 u_S^2}\right]\, .
\ee
This agrees with the entropy function computed in
\cite{0508042}. 

The  relations \refb{e2.11} between the two sets of charges
depend on the precise normalization of the dual
field $\HH^{(6)}_{MNP}$ and the definition of the four dimensional
gauge fields in terms of the six dimensional fields, but not on the
details of the Lagrangian density $\LL^{(6)}$. Thus \refb{e2.11}
continues to hold even after inclusion of higher derivative corrections
to the action.
In order
to facilitate comparison with the known results we shall express
all answers in terms of 
the charges $q_1$, $q_3$, $p_2$ and $p_4$ from now on.
Physically these charges represent $n$ unit of momentum and $w$
unit of winding charge along $x^4$ and $N'$ unit of Kaluza-Klein
monopole and $W'$ unit of H-monopole charge associated with
the circle along $x^5$, with\cite{0508042}
\be{eqw}
q_1={1\over 2}\, n, \qquad q_3 = {1\over 2}\, w, \qquad
p_2=4\pi N', \qquad p_4 = 4\pi W'\, .
\ee

Extremizing \refb{e2.10a} with respect to $v_1$, $v_2$,
$u_1$, $u_2$ and $u_S$ and using \refb{e2.9} we get
\bea{e2.12}
&& v_1 = v_2 ={1\over 4\pi^2} \,\left| p_2 p_4\right|, \quad
u_S=8\pi \sqrt{\left| q_1 q_3\over p_2 p_4\right|}, \quad
u_1=\sqrt{\left|q_1\over q_3\right|}, \quad u_2 = \sqrt{\left|
p_4\over p_2\right|} \cr
&& \wt e_1 = {1\over 4\pi q_1} \, \sqrt{|p_2 p_4 q_1 q_3|}, \qquad
\wt e_3 = -{1\over 4\pi p_4}\, \sqrt{|p_2 p_4 q_1 q_3|}
\, .
\eea
Substituting this back into \refb{e2.10a} we get the leading order
contribution to the black hole entropy:
\be{e2.12a}
\EE_0= \sqrt{|p_2 p_4 q_1 q_3|}=2\pi\sqrt{|nwN'W'|}\, .
\ee

We now turn to the evaluation of $\EE_1$.  We shall divide the
contribution into two parts:
\be{e2.13}
\EE_1=\EE_1'+\EE_1''\, ,
\ee
where
\be{e2.14}
\EE_1' = -2\pi   \int d\theta d\phi dx^4 dx^5\,  
\sqrt{-\det G^{(6)}}\, 
\LL^{(6)}_1
\ee
and
\be{e2.15}
\EE_1''= - \int d\theta d\phi
\sqrt{-\det G}
\wt\LL'' \, .
\ee
First let us compute $\EE_1'$. For this we need the expression for the
four derivative corrections to the heterotic string effective action at 
the string tree level. This is given by\cite{tseyt,hull}
\bea{e2.2}
\LL^{(6)}_1&=&{1\over 16\pi} \, e^{-2\Phi^{(6)}}\, \Bigg[ 
R^{(6)}_{KLMN}
R^{(6)KLMN} -{1\over 2} R^{(6)}_{KLMN} H_P^{(6)KL}
H^{(6)PMN}  \cr
&& -{1\over 8} H_K^{(6)MN} H^{(6)}_{LMN}H^{(6)KPQ}
H^{(6)L}_{PQ} + {1\over 24} H^{(6)}_{KLM}
H^{(6)K}_{PQ} H_R^{(6)LP} H^{(6)RMQ} 
\Bigg]\, . \nonumber \\
\eea
Using \refb{e2.5}-\refb{e2.7} and \refb{e2.14} we  get\cite{0607094}
\bea{e2.16}
\EE'_1 &=& -{4\pi } v_1 v_2 u_S
\Bigg[ {1\over 2 v_1^2} + {1\over 2 
v_2^2} -
{3 \wt e_1^2 u_1^2\over v_1^3} - {3 \wt p_2^2 u_2^2\over 16 v_2^3 \pi^2}
+ {11 u_1^4 \wt e_1^4\over 2 v_1^4} + {11 \wt 
p_2^4 u_2^4\over 512 v_2^4 
\pi^4} \nonumber \\
&& - {4 u_1^2 \wt p_4^2 \over v_1 v_2^2 u_S^2} -
{64 \pi^2 u_2^2 \wt e_3^2\over v_1^2 v_2 u_S^2}
+ {4 u_1^4 \wt p_4^2\wt e_1^2 \over v_1^2 v_2^2 u_S^2}
+ {4 u_2^4 \wt e_3^2 \wt p_2^2 \over v_1^2 v_2^2 u_S^2}
\nonumber \\
&& - {40 u_1^4 \wt p_4^4\over v_2^4 u_S^4} -
{10240\, \pi^4 u_2^4 \wt e_3^4 \over v_1^4 u_S^4}\Bigg]\, .
\eea
As discussed below
eq.\refb{ee1}, in computing the black hole entropy we can substitute
the solution given in  \refb{e2.12}, 
obtained by extremizing
$\EE_0$, into the expression for $\EE_1$. This gives the contribution
to the black hole entropy from $\EE_1'$ to be\cite{0607094}
\be{e2.17}
\EE_1' = 16\pi^2 \, \sqrt{\left|  {q_1 q_3\over p_2 p_4}\right| } \, .
\ee

Let us now turn to the computation of $\EE_1''$. 
This would require
first dimensionally reducing the  
Chern-Simons term to construct a
covariant four dimensional Lagrangian density 
via eq.\refb{ed1b}, and then computing its contribution to the
entropy function via eq.\refb{e2.15}. 
This analysis can be simplified
by regarding the sphere labelled by $\theta$, $\phi$ also
as a compact space and considering dimensional reduction
of \refb{ed1b} all the way to two dimensions spanned 
by the coordinates
$r$ and $t$.
The resulting two dimensional Lagrangian density has the form
\bea{e2.18}
\sqrt{-\det G^{(2)}} \, \wt\LL^{(2)\prime\prime}
&=&  - {1\over 16\pi^2} {1\over (3!)^2}
\lambda \, \int dx^4 dx^5 \, d\theta \, d\phi\, 
\epsilon^{MNPQRS} \HH^{(6)}_{MNP} 
\Omega^{(6)}_{QRS} \nonumber \\
&& + \hbox{total derivative terms}\, ,
\eea
where the total derivative terms need to be chosen such that the
$\LL^{(2)\prime\prime}$ is manifestly covariant. The contribution
$\EE_1''$ to the entropy function is then given by
\be{e2.19}
\EE_1'' = -2\pi \, \sqrt{-\det G^{(2)}} \, \wt\LL^{(2)\prime\prime}\, ,
\ee
evaluated in the near horizon background of the black hole.

We can carry out the dimensional reduction from six to two
dimensions in two statges. First of all we note that the six
dimensional field
configuration given in \refb{e2.5} has the structure of a product
of two three dimensional spaces, the first one labelled by
$(\theta,\phi,x^5)$ and the second one labelled by $(t,r,x^4)$.
Thus we can make a consistent truncation where we consider
only those field configurations which respect this product structure.
In this case \refb{e2.18} simplifies to
\be{e2.20}
\sqrt{-\det G^{(2)}} \, \wt\LL^{(2)\prime\prime}
= - {1\over 16\pi^2} {1\over (3!)^2}
\lambda \, \int \, dx^4\, dx^5 \, d\theta \, d\phi\, 
\epsilon^{\wm \wn \wpp} \epsilon^{\wa\wbb\wg}
(\HH^{(6)}_{\wm \wn \wpp} 
\Omega^{(6)}_{\wa\wbb\wg} - \Omega^{(6)}_{\wm \wn \wpp} 
\HH^{(6)}_{\wa\wbb\wg})
\ee
where the indices $\wm$, $\wn$, $\wpp$ run over $(\theta,\phi,x^5)$
and
the indices $\wa$, $\wbb$, $\wg$ run over $(t,r,x^4)$.
We have chosen the following convention for the three
dimensional $\epsilon$ tensors:
\be{e2.20a}
\epsilon^{tr4}=1, \qquad \epsilon^{\theta\phi 5} = 1\, .
\ee
Let us now label the components of the six dimensional metric
as
\be{e2.21}
G^{(6)}_{\wm\wn}dx^{\wm}dx^{\wn} = 
G^{(6)}_{55} \left( h_{mn} dx^m dx^n 
+ (d x^5 + 2 \AAA^{(2)}_m dx^m)^2\right)  
\ee
and
\be{e2.22}
G^{(6)}_{\wa\wbb}dx^{\wa}dx^{\wbb} = 
G^{(6)}_{44} \left( g_{\alpha\beta} dx^\alpha dx^\beta 
+ (d x^4 + 2 \AAA^{(1)}_\alpha dx^\alpha)^2\right)  
\ee
where the indices $m,n$ run over $(\theta,\phi)$ and the indices
$\alpha,\beta$ run over $(t,r)$. Then it follows from
the analysis of \cite{0305117,0601228} that
\be{e2.23}
\int \, dx^5 \, d\theta \, d\phi\, \epsilon^{\wm \wn \wpp}
\Omega^{(6)}_{\wm \wn \wpp}
= 4\pi \, \int d\theta \, d\phi\, \epsilon^{mn}
\left[ R_h \, \FF^{(2)}_{mn} + 4 \, h^{m'p'}\, 
h^{q'q}\, \FF^{(2)}_{mm'}\, \FF^{(2)}_{p'q'}
\, \FF^{(2)}_{qn}\right]
\ee
and
\be{e2.24}
\int \, dx^4  \epsilon^{\wa \wbb \wg}
\Omega^{(6)}_{\wa \wbb \wg}
= 4\pi \,  \epsilon^{\alpha\beta}
\left[ R_g \FF^{(1)}_{\alpha\beta} + 4 \, g^{\alpha'\gamma'}\, 
g^{\delta'\delta}\, \FF^{(1)}_{\alpha\alpha'}
\FF^{(1)}_{\gamma'\delta'}\, \FF^{(1)}_{\delta\beta}\right]
+ \hbox{total derivative terms}
\ee
where $R_h$ and $R_g$ denotes the scalar curvature associated with
the metrics $h_{mn}$ and $g_{\alpha\beta}$ respectively.
Our convention for the two dimensional $\epsilon$ tensor
is
\be{e2.25a}
\epsilon^{tr}=1, \qquad \epsilon^{\theta\phi}=1\, .
\ee
Thus we get
\bea{e2.25}
&& \sqrt{-\det G^{(2)}} \, \wt\LL^{(2)\prime\prime}
\cr
&=&  - {1\over  \pi} {1\over (3!)^2}
\lambda \, \Bigg[ \, 6\pi \, \left(\int  \, d\theta \, d\phi\, 
\epsilon^{mn} \,
\HH^{(6)}_{5mn} \right) \,  \epsilon^{\alpha\beta}
\left[ R_g \FF^{(1)}_{\alpha\beta} + 4 \, g^{\alpha'\gamma'}\, 
g^{\delta'\delta}\, \FF^{(1)}_{\alpha\alpha'}
\FF^{(1)}_{\gamma'\delta'}\, \FF^{(1)}_{\delta\beta}\right]\cr
&& - 6\pi \, \left( \int d\theta \, d\phi\, \epsilon^{mn}
\left[ R_h \, \FF^{(2)}_{mn} + 4 \, h^{m'p'}\, 
h^{q'q}\, \FF^{(2)}_{mm'}\, \FF^{(2)}_{p'q'}
\, \FF^{(2)}_{qn}\right] \right) \, \epsilon^{\alpha\beta}
\HH^{(6)}_{4\alpha\beta} \Bigg]\, .
\eea

Since the lagrangian density now has manifest covariance, we can
apply the entropy function formalism.
This requires evaluating the right hand side of \refb{e2.25} for the
six dimensional background given in \refb{e2.5}. Noting that
for this configuration
\be{e2.26}
h_{mn} dx^m dx^n = v_2\, u_2^{-2} \, 
(d\theta^2 +\sin^2\theta d\phi^2)\, , \qquad
g_{\alpha\beta} dx^\alpha dx^\beta =
v_1 \, u_1^{-2}\, (-r^2 dt^2 + dr^2/r^2)\, ,
\ee
we get
\be{e2.27}
\sqrt{-\det G^{(2)}} \, \wt\LL^{(2)\prime\prime}
= {2\lambda \pi\over 3} \, \Bigg[ {\wt p_4\over 4\pi}
\left( {u_1^2 \over v_1}\wt e_1 - 2 {u_1^4\over v_1^2}\,
\wt e_1^3\right) + \wt e_3 \, \left( {u_2^2\over v_2} \, {\wt p_2\over
4\pi} - 2 {u_2^4 \over v_2^2} \, \left( {\wt p_2\over 4\pi}\right)^3
\right)
\Bigg]\, .
\ee
Evaluating this for the solution given in \refb{e2.12} we get
\be{e2.28}
\EE_1'' = -2\pi \sqrt{-\det G^{(2)}} \, \wt\LL^{(2)\prime\prime}
= {1\over 6}\, \lambda\, \pi^2 \left( 
{q_1 q_3\over \sqrt{\left|p_2p_4q_1q_3\right|}} 
+ {\sqrt{\left|p_2p_4q_1q_3\right|}
\over p_2 p_4}\right)
\ee
For definiteness we shall now consider the range of values
\be{e2.29}
p_2>0, \qquad p_4>0, \qquad q_3> 0\, .
\ee
In this case the full black hole entropy, given by the value
of the entropy function at its extremum, becomes
\be{e2.31}
\EE =  \EE_0 + \EE_1' + \EE_1''
= \sqrt{|p_2 p_4 q_1 q_3|} \left[ 1 + {\pi^2\over p_2 p_4} \left\{16 
+ 
{1\over 6}\, \lambda\, \left(1 + {q_1\over |q_1|}\right) 
\right\}\right]\, .
\ee

Let us now turn to the determination of the parameter $\lambda$.
If we define
\be{eax1}
a = 128\pi \, \BB^{(6)}_{45}\, ,
\ee 
then after elimination of $H^{(6)}_{MNP}$ using 
\refb{e2.3} and dimensional reduction to four dimensions,
the action contains the terms:
\be{eax2}
{1\over 32\pi} \, \int \, d^4 x\, 
\left[ -{1\over 2} \, \sqrt{-\det G} \, e^{2\Phi}\, 
G^{\mu\nu}\p_\mu a \p_\nu a
+ {\lambda\over 48}\, a \, \epsilon^{\mu\nu\rho\sigma}
\, R^c_{~d\mu\nu} \, R^d_{~c\rho\sigma}+\ldots\right]\, .
\ee
$a$ plays the role of the axion field.
Comparing this with the standard action for tree level
heterotic string theory (see {\it e.g.} \cite{0603149})
compactified down to four dimensions, we get
\be{eax3}
\lambda= 48\, .
\ee

Eq.\refb{e2.31} now gives
\bea{e2.34}
\EE &=& \sqrt{|p_2 p_4 q_1 q_3|} 
\left[ 1 + 32\, {\pi^2\over p_2 p_4}\right] 
= 2\pi\sqrt{|nwN'W'|} \left[ 1 + {2\over N'W'}
\right] \qquad \hbox{for $q_1>0$}\, ,
\cr
&=& \sqrt{|p_2 p_4 q_1 q_3|} 
\left[ 1 +  16 {\pi^2\over p_2 p_4}\right] 
= 2\pi\sqrt{|nwN'W'|} \left[ 1 + {1\over N'W'}
\right] \qquad \hbox{for $q_1<0$}\, .
\nonumber \\
\eea
For $q_1>0$ the black hole is  supersymmetric. The result for the
entropy 
agrees with the result obtained by 
1) including only the  Gauss-Bonnet term in the four
dimensional effective action\cite{9711053,0508042},
2) including  a fully supersymmetrized version of the
curvature squared correction in the four dimensional effective 
action\cite{9812082,9904005,9906094,9910179,0007195,
0009234,0012232}
and 3) the argument based on the existence of an $AdS_3$ component
of the near horizon geometry and supersymmetry of the associated
boundary theory\cite{0506176,0508218}. 
Since the last result makes use of  supersymmetry 
to relate the gauge anomaly to the trace anomaly in the
boundary theory, our result provides an indirect evidence that the
bosonic effective action given in \refb{e2.2} can be consistently
supersymmetrized to this order in $\alpha'$.

We also see from 
\refb{e2.31} that
\be{e2.32}
\EE_{q_1>0} - \EE_{q_1<0} = 16 \sqrt{|p_2 p_4 q_1 q_3|}\, 
{\pi^2\over p_2 p_4}\, .
\ee
This agrees with the result derived under the assumption 
that the subspace spanned by the coordinates $x^4$, $t$ and $r$
form a locally $AdS_3$ space time near the 
horizon\cite{0506176,0508218,0603149,0601228}.

Finally we note that for heterotic string theory compactified on $T^6$
or more general $\NN=4$ supersymmetric string compactification,
the statistical entropy of some of
these black holes can be computed exactly
by representing them as a configuration of D-branes and Kaluza-Klein
monopoles in the dual type IIA string 
theory\cite{9607026,0505094,0506249,
0510147,0602254,0603066,0605210}.
The approximation used here by restricting to tree level heterotic
string theory will be a valid approximation if the near horizon
value of the string coupling constant is small.  
\refb{e2.12} shows that this
requires the electric charges $q_1$, $q_3$ to be
large compared to the
magnetic charges $p_2$, $p_4$. Within this approximation the result
for the statistical entropy is known to agree with the black hole
entropy computed using the Gauss-Bonnet
term\cite{0412287,0510147,0602254,0605210}. 
Hence this also agrees
with the results found here by including the complete set of higher
derivative terms.

 \end{document}